**Dirk Helbing (ETH Zurich)**

# FuturICT – New Science and Technology to Manage Our Complex, Strongly Connected World

**We have built particle accelerators to understand the forces that make up our physical world. But we still don't understand the principles underlying our strongly connected, techno-socio-economic systems. To fill the knowledge gaps and keep up with the fast pace at which our world is changing, a *Knowledge Accelerator* must urgently be created. An interdisciplinary integration of natural, social, and engineering sciences, as it will be performed by the FuturICT flagship project, can produce the synergy effects required to address many of our 21$^{st}$ century challenges. One can expect that, after the age of physical, biological and technological innovations, Europe can lead the next era – a wave of social and socio-inspired innovations.**

*Globalization and technological change have made our world a different place. This has created or intensified a number of serious problems, such as international conflict and world-wide terrorism, global financial and economic crises, political instabilities and revolutions, the quick spreading of diseases, disruptions of international supply chains, organized crime and increased cyber-risks. However, there are also new opportunities to create unprecedented benefits for our economy and society, based on a whole range of new methods and innovations.*

*Although creating more interconnected systems has contributed to the above problems, future information and communication technologies (ICTs) can also be key to the solution. This requires us to establish a new science of multi-level complex systems, bringing the best knowledge of experts on information and communication systems, complex systems and the social sciences together. The FuturICT flagship project will[a] develop the capacity to explore and manage our future, based on a fundamental understanding of the institutional and interaction-based principles that make connected systems work well.*

*The methods and Big Data needed for such a scientific endeavour are now becoming available: it is, therefore, time to make a federated and open Big Science effort akin the Human Genome Project. Open, because we need to prevent private monopolies of socio-economic data; federated, because joint efforts are the only way to tackle humanity's global challenges and ensure leadership in socio-inspired ICT innovations. The investments into the FuturICT's project can benefit citizens and society multiple times, by mitigating global problems and systemic risks, and by creating new possibilities to participate in social, economic and political affairs. In particular, FuturICT will create new spin-offs, business opportunities and jobs.*

**The Need for New Knowledge in a Fundamentally Changed World**

Today, neither past knowledge nor established policies seem sufficient anymore to manage the future (see Box 1). This is because technological, social and economic systems are becoming more and more complex, and also mutually interdependent. Such strongly connected systems often behave completely different from loosely connected systems and to what our everyday intuition suggests, a situation raising fundamental scientific challenges, but also ethical ones (see Box 3):

- The dynamics of strongly connected systems with positive feedbacks is faster.
- Extreme events occur more often and can impact the whole system.
- Self-organization and strong correlations dominate the dynamics of the system.
- The system behaviour is often counter-intuitive, and unwanted feedback or side effects are typical.
- The system behaviour is hard to predict, and planning for the future may not be possible.
- Opportunities for external control are very limited.
- Even the most powerful computers cannot perform an optimization of the system behaviour in real time, as the number of interacting system elements is too large.
- The competition for limited resources implies reduced redundancies in the system and a larger vulnerability to random failures or external shocks.
- The loss of predictability and control lead to an erosion of trust in private and public institutions, which in turn can lead to social, political, or economic destabilization.

Such strongly coupled systems cannot be managed well in a top-down fashion. Rather than controlling the individual elements of a system, it becomes crucial to stimulate a more favourable self-organization in the whole system by establishing suitable interaction rules (the `rules of the game') (see Table 1). Bottom-up elements allow for greater flexibility, efficiency, and resilience of the system (see Box 5).

**Future ICT Systems as Artificial Social Systems**

Our ICT systems are increasingly suffering from similar problems that worry societies: the lack of coordination, instability, an inefficient use of resources, conflicts of interest. The recent explosion of cyber-crime and the new notion of cyber-war leave the impression that conventionally operated ICT networks may get out of control. This is happening because ICT systems are usually not tested for the systemic interactions of their components. Yet, they are complex systems, which are made up of billions of non-linearly interacting elements (computers, smartphones, software agents etc.). More and more, these components take autonomous decisions based on an internal representation ("subjective" interpretation) of the surrounding world and expectations regarding future conditions. This effectively makes them *artificial* *social systems*. For example, computer-based automatic trading strategies now perform the majority of transactions in our world's financial system.

| Our Previous World | Our New World |
|---|---|
| Separate or weakly connected system components, local or regional interactions | Strongly connected and interdependent system components, global interactions |
| Dominated by the (visible) system components | Dominated by their (invisible) interactions |
| Simple system behaviour | Complex system behaviour |
| Sum of properties of individual system components characterizes system behaviour | Emergent collective behaviour, implies new (and often unexpected) system behaviour |
| Conventional wisdom works well | Counter-intuitive behaviour, extreme events are common |
| Well predictable and controllable in top-down fashion | Less predictable, management by setting rules for bottom-up self-organization |

**Table 1:** *Comparison of a weakly connected world with local or regional interactions (as we tended to have it in the past) with a strongly connected world (as we are facing it today). In order to understand systemic risks resulting from the new interdependencies and to develop the ability of integrated risks management, we need to overcome the classical silo thinking and even more than that: We need a new way of thinking, a paradigm shift from focusing on the components of a system to focusing on their non-linear interactions, as studied by complexity science. This paradigm shift is of similar importance as the one from a geocentric to a heliocentric worldview, triggered by Galileo. It will create a new understanding of our techno-socio-economic-environmental system, and facilitate new solutions to long-standing problems.*

Currently, most information and communication systems are not designed for the collective behaviour that may result from the interaction of their components. (The same is also true for socio-economic systems.) As we go on connecting these systems more and more densely, this will bring about a lack of robustness (failure tolerance) and a lack of resilience (i.e. a vulnerability to attacks and external shocks). Given the ubiquitous use of ICT systems and our strong dependence on the reliability of these systems, proper design principles for such socially interactive systems must be urgently identified. This ultimately requires fundamental knowledge from the social sciences.

**The Urgent Need of a Federated Big Science Approach**

The complexity in our ICT systems and the techno-socio-economic challenges of humanity in the 21$^{st}$ century require our society to make a large-scale federated investment to fill the current knowledge gaps (see Box 7). The FET Flagship Call by the European Commission provides a unique and timely opportunity for this (see Box 8). We urgently need to learn how to manage our future in a complex, strongly connected world. In fact, the 30 years delay in developing realistic scientific models of our globalized techno-socio-economic world and the lack of a coordinated approach to assess systemic risks and develop an integrated risk management are alarming.

**Need for a New Multi-Level Complex Systems Science**

The challenges of the 21$^{st}$ century require the development of a new kind of complexity science: the science of *multi-level complex systems* focusing on realistic models rather than just metaphorical analyses. This new science should allow us to understand not only the impact of a system component on others, but also the resulting links between micro-level interactions and macro-level behaviour (and vice versa). We also need to understand the complex interdependencies between the different institutions, infrastructures and networks on which our society is built. This requires social scientists to ask the right questions and provide a characterization of the system components (individuals, institutions, etc.). It requires complexity scientists to gain theoretical insights by studying systemic interactions of these components. And it requires computer scientists and information and communication experts to create methods, data and platforms that will allow us to understand and manage our world better (see Figure 1).

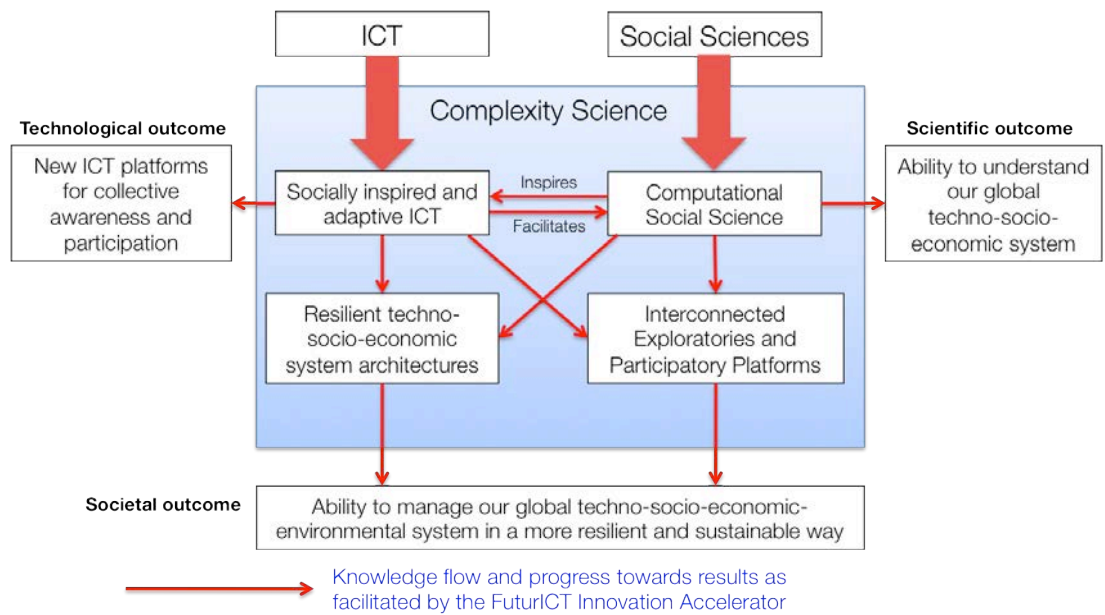

**Figure 1.** *The interdisciplinary concept of FuturICT foresees the integration of expertise in information and communication technology (ICT), complexity and social sciences.*

**Need for a New Data Science**

This calls for a new Data Science (or *Social Information Theory),* which focuses not just on bits and bytes, but also on the meaning and impact of information (as we need to understand under what conditions and how new knowledge is created from existing pieces of information). It also requires a considerably extended complexity science, which studies not only the stylized patterns and dynamics resulting from the non-linear interaction of *simple* elements. It also needs to understand the result of interactions between individuals with cognitive complexity or system elements with a complex response to the surrounding world. Such systems with various levels of complexity are probably not anymore analytically tractable and, therefore, require the use of future supercomputers.

Despite the urgent need for such a multi-level complex systems science and although many of its components have been created in the past, a coordinated effort could not yet take off due to institutional obstacles and a lack of resources. The FuturICT project will, for the first time, integrate all necessary competencies by bringing together the best of all available knowledge from the engineering, natural and social sciences (see Figure 2).

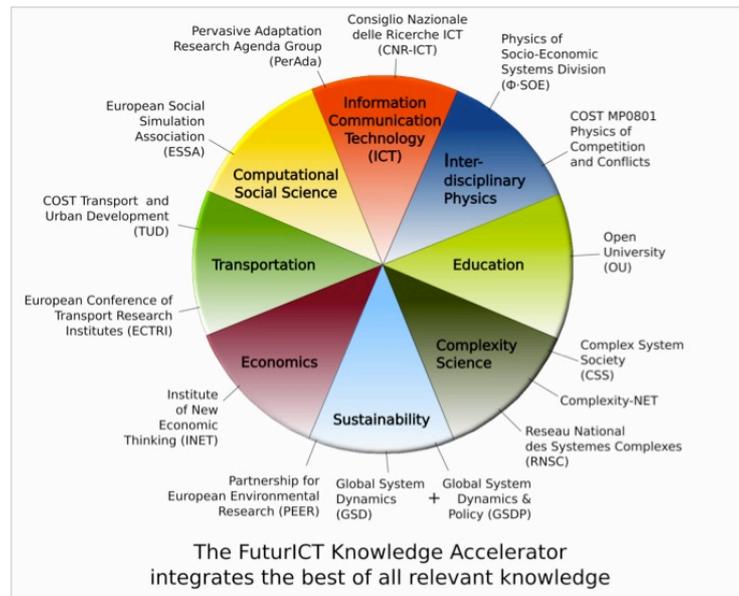

**Figure 2.** *FuturICT is a multi-disciplinary project integrating many different research areas. This is done with the support of various established science organizations, which stay in close contact with their respective research communities.*

This new science will be boosted by the availability of vast amounts of data from a wide variety of techno-socio-economic systems. In fact, future sensor networks will produce more live stream data than can be stored or moved around. To make use of them, they must be aggregated "on the fly" - and in a privacy-respecting way (see Box 1). But these data will also make it possible to create something like a "*Planetary Nervous System"* and, thereby, to create collective (self-)awareness of the impact that our decisions and actions are likely to have on our techno-socio-economic-environmental system (see Box 10).

In turn, an ability to quantify the social impact of our actions will help us to avoid decisions that exploit or destroy the socio-economic fabric on which our society is built, for example, social capital, solidarity and trust. It will eventually promote a more responsible behaviour, just as the measurement of the environmental footprint has done. Developing the ability to quantify the *social footprint* seems a particularly promising way to successfully establish *sustainable systems.*

## What FuturICT Will Do

*A Way Forward, Aided by Information*

The complexity of modern technology lies far beyond the capacity of the human brain to comprehend or analyse in detail. Information technology can considerably expand this capacity. For example, scientists and engineers rely on massive computer power and data processing to design and test everything from cars and electronic devices to medical drugs. We face even greater complexity in our socio-economic systems, especially in the interaction with the rapidly expanding technological infrastructures such as the Internet and the Earth's vast, multi-component environment. Only recently, however, have we begun to exploit the power of information technology to gain a better understanding of the human-Earth system, and to improve our capacity to manage this system on the basis of well-founded knowledge.

The FuturICT project aims to develop new science and technology, capitalizing on the current data revolution. The project will develop a visionary information platform, considering insights from ethics, complexity and social sciences. This system will be able to act as a *Flight Simulator* or *Policy Wind Tunnel*, allowing people to test multiple options in on a complex and uncertain world, and pluralistic perspectives on it. The platform would analyse data on a massive scale and leverage them with scientific knowledge, thereby giving politicians, decision-makers and citizens a better basis to base their decision on. In perspective, this would enable everyone of us to explore the possible or likely consequences of even barely imaginable scenarios, effectively helping us to see just a little around the corner into possible futures (see Box 11).

The potential benefits are huge: reducing the impact of major societal and economic problems by only 1 percent would save the European Union billions of Euros every year (see Box 12). Indeed, the social and complexity sciences can present a number of recent impressive success stories (see Box 2). Thus, similar to weather forecasts, it is expected that FuturICT can create value that is many times higher than the required investments.

*How it Can Happen*

To achieve its goals, the FuturICT project will develop new information and communication technology (ICT) to collect massive data sets and mine them for useful or meaningful information. It will also have the capacity to self-organize and adapt to the collective needs of users. These ICT systems will be the basis of the *FuturICT Platform* (or: *Living Earth Platform)* (see Figure 3). It will be built on three new inter-connected instruments to gain novel insights into our world: the *Living Earth Simulator*, the *Planetary Nervous System* and the *Global Participatory Platform* (see Figure 4).

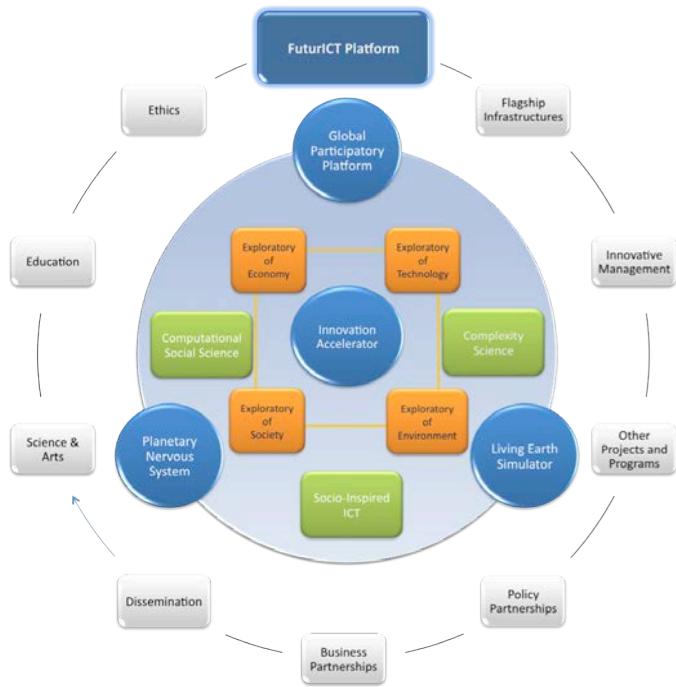

**Figure 3.** *Main components and activity areas of the FuturICT flagship project.*

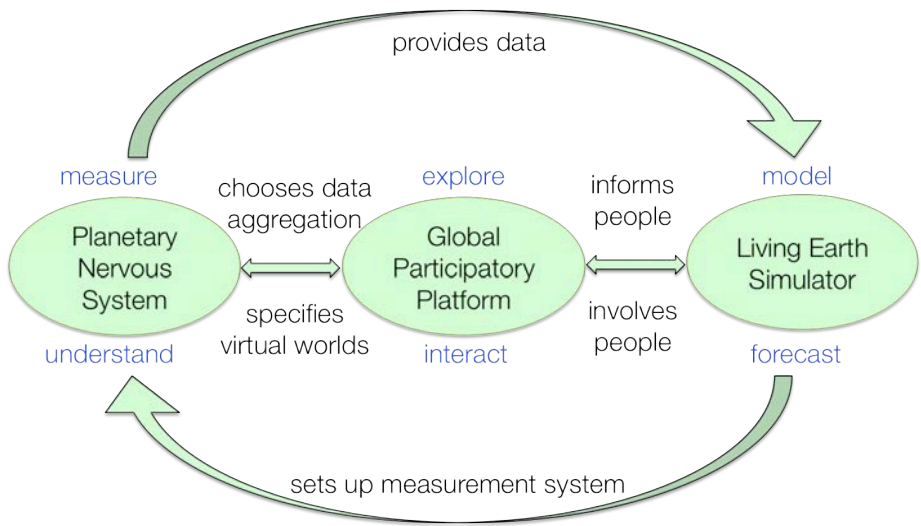

**Figure 4.** *Interdependencies between FuturICT's main ICT components. The project will develop 3 key methods and instruments to study our techno-socio-economic-environmental world: The Living Earth Simulator, the Planetary Nervous System, and the Global Participatory Platform. Analogously to microscopes and telescopes, they may be considered to establish the Socioscopes of the future.*

The *Living Earth Simulator* will enable the exploration of future scenarios at different degrees of detail, employing a variety of perspectives and methods (such as sophisticated agent-based simulations and multi-level models). Exploration will be supported via an envisaged *World of Modelling* – an open software platform, comparable to an app-store, to which scientists and developers can upload theoretically informed and empirically validated modelling components that map parts of our real world. The Living Earth Simulator will require the development of interactive, decentralized, scalable computing infrastructures, coupled with an access to huge amounts of data, which will become available by integrating various data sources coming from online surveys, web and lab experiments, and from large-scale data mining.

This is where the *Planetary Nervous System* comes in. It can be imagined as a global sensor network, where "sensors" include anything able to provide data in real-time about socio-economic, environmental or technological systems (including the Internet). Such an infrastructure will enable real-time data mining ("reality mining"), and the calibration and validation of coupled models of socio-economic, technological and environmental systems with their complex interactions. It will even be possible to extract suitable models in a data-driven way, guided by theoretical knowledge.

The *Global Participatory Platform* will promote communication, coordination, cooperation and the social, economic and political participation of citizens beyond what is possible through the eGovernance platforms of today. In this way, FuturICT will create opportunities to reduce the gap between users and providers, customers and producers etc., facilitating a participation in industrial and social value generation chains. Building on the success principles of Wikipedia and the Web2.0, societies will be able to harness the knowledge and creativity of multiple minds much better than we can do today. The Global Participatory Platform will also support the creation of *Interactive Virtual Worlds*. Using techniques such as serious multi-player online games, we will be able to explore possible futures – not only for different designs of shopping malls, airports, or city centres, but also for different financial architectures or voting systems.

In addition to the interconnected systems forming the Living Earth Platform, FuturICT will also create an *Innovation Accelerator* that will identify innovations early on, distil valuable knowledge from a flood of information, find the best experts for projects, and fuel distributed knowledge generation through modern crowd sourcing approaches. In particular, the Innovation Accelerator will support communication and flexible coordination in large-scale projects, co-creation, and quality assessment. Hence, the Innovation Accelerator will also form the basis of the innovative management of the FuturICT flagship. Beyond this, it will fill the vision of Europe's Innovation Union with life and create many new business opportunities, e.g. based on socio-inspired innovations (see Box 13).

*It is Time for Practical Steps to a Better Future*

To succeed with its ambitious endeavour, the FuturICT project team is building communities in most European countries and other continents, bridging between ICT, social and complexity sciences. It will build the FuturICT platform by integrating 4 interconnected Exploratories, which are ICT infrastructures to explore our global techno-socio-economic-environmental system through the combination of large-scale data mining, multi-level modelling, supercomputing and participatory approaches.

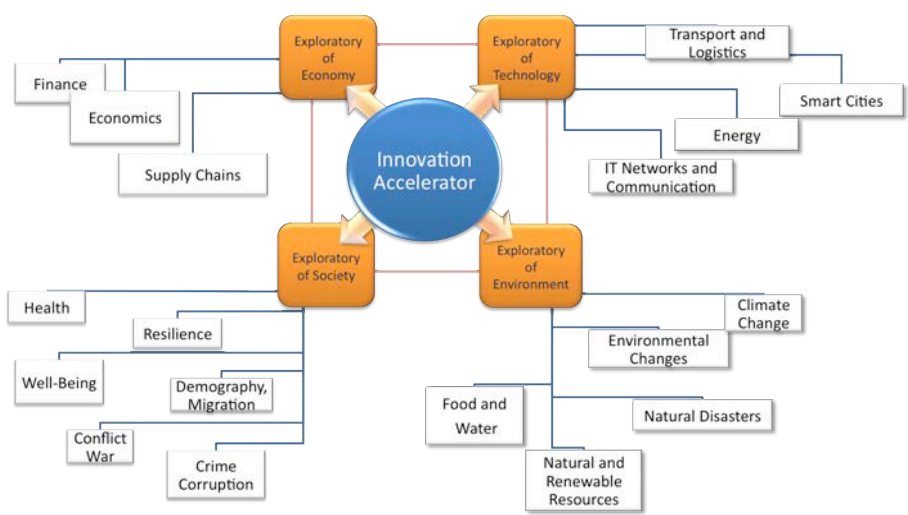

**Figure 5.** *The Exploratories of Society, Economy, Technology, and Environment are established by integration of various Observatories. The Exploratory of Society, for example, could be composed of Observatories for Social Well-Being, Health, Crime and Corruption, Conflict and War, Demography and Migration, and Resilience.*

The Exploratories of Society, Economy, Technology, and Environment are composed of various interactive Observatories, for example, Observatories of Financial and Economic Systems, of Conflicts and Wars, of Social Well-Being, of Health Risks, of Transportation and Logistics etc. (see Figure 5). Integrating the Observatories and Exploratories over time (see Figure 6) will overcome "disciplinary silo thinking" and eventually facilitate a systemic picture of risks and opportunities as well as integrated risk management.

Note that FuturICT will pursue a *pluralistic* approach, allowing one to study many different perspectives in parallel. This will provide a more differentiated picture of the interactions on our planet and allow us to better manage our way forward in a rapidly changing world.

*Potential and Need of Socio-Inspired Technologies*

FuturICT's research program will also be crucial for the effective design of future ICT systems, since these are indeed becoming socially interactive systems (see the Section on *Artificial Social Systems* above). As our society is now largely dependent on information and communication technologies, their stability and reliability has become absolutely crucial – but at least for current designs, this stability is not guaranteed. Systemic breakdowns, cyber-crime and cyber-war are problems that have recently become virulent and show the vulnerability of the systems to cascading failures and other problems (see Box 6). At the same time, several social features such as self-organization, adaptiveness, emergent cooperation, social norms, cultures and community formation are new attractive features of future ICT systems. *Trust* is just one example of a hardly understood, but crucial property of our social and ICT systems. The creation of a *Trustable Web,* based on principles of social, reputation-based self-control in order to keep cyber-crime down, is probably the most important example of a future socio-inspired ICT system.

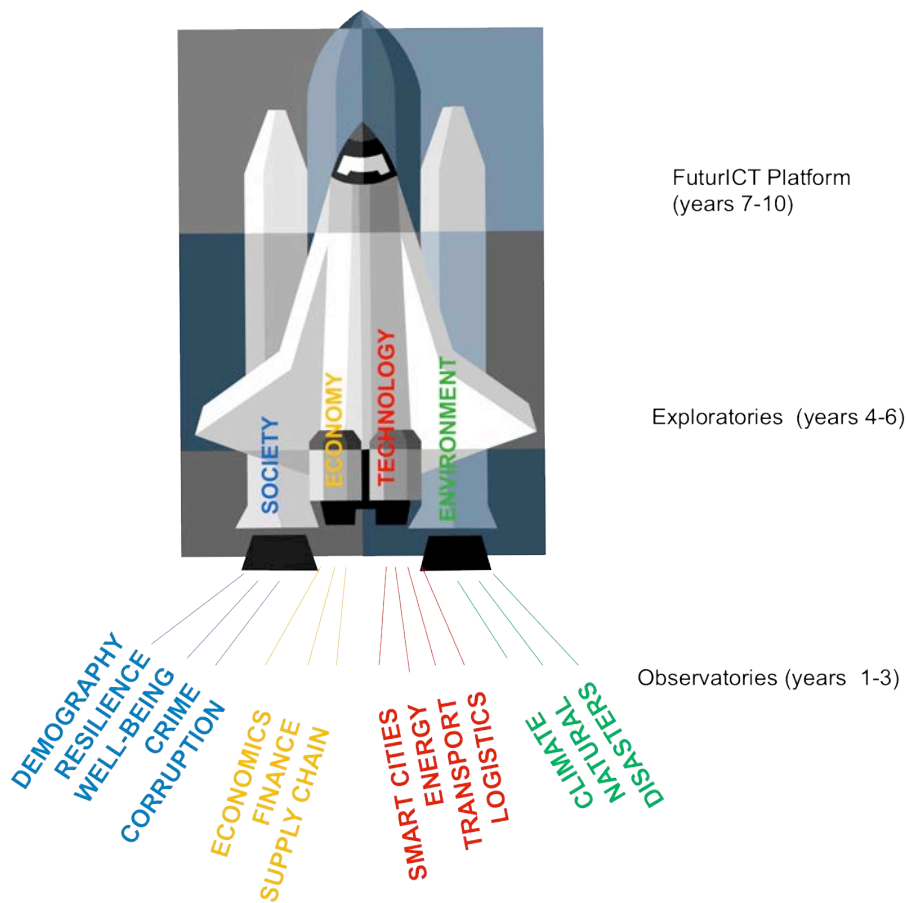

**Figure 6.** *The FuturICT platform is built by integrating various Observatories into closely interconnected Exploratories of Society, Economy, Technology, and Environment, and progressively taking into account the interactions between them.*

## What FuturICT Will NOT Do

Most importantly, the FuturICT project will NOT attempt to collect "all the data in the world", or to represent each individual on the globe by an identical copy in some giant multi-agent simulation, considering private features and preferences of all individuals. *Science is the art of abstraction and approximation.* Just as maps do not show all the features of our environment, a scientific model is specified such that a particular question can be addressed in the simplest possible way. That is, models are to be problem-specific, and parameters and variables not expected to be relevant for the answer should be neglected. In many cases, one is interested in global interdependencies on the aggregate level. Then, a macroscopic description is sufficient. Most computer simulations are based on a multi-level approach and not on the micro-simulation of the individual system elements.

As a matter of principle, it would obviously not be possible or desirable to simulate each individual in detail, considering the complexity of its cognitive dynamics. However, the interaction of many system elements often reduces the dimensionality of the relevant system dynamics (i.e. there is a small fraction of variables that matter, while many variables do not change over the relevant time scale and others change so quickly that they may be treated as random variables). In other words, the largest amount of complexity probably occurs on the 'micro-level' of the individual, while collective behaviour tends to be simpler due to many factors such as herding effects, social conventions, norms, and laws (otherwise we would not have cultural trends, fashions etc.). This implies the chance of probabilistic short-term forecasts similar to weather forecasts (see Box 11). These are sufficient for adaptive management approaches, which can reach considerable improvements (see Box 5).

**Organizational Principles**

Setting funding issues aside, the strategy of the partners of the currently running FuturICT Pilot project is to formulate a visionary goal, to elaborate a project with an Apollo-level ambition (as expected by the FET flagship program), to identify the related grand scientific challenges, to develop a research strategy and roadmap, to form an integrated multi-disciplinary community, and to develop a platform for global collaboration and exchange.

By now, the FuturICT project is supported by a large and quickly growing multi-disciplinary community (see Box 15). FuturICT enjoys the support of many European countries and also of a strong US and Japanese community. Step by step, the project is creating links to China, Singapore, Australia, South-American and African countries. In several of these countries, there is a strong desire to participate in addressing the global challenges FuturICT will tackle, based on complementary national budgets. Therefore, FuturICT is trying to create an open platform with interfaces that would allow other countries and projects to team up. One may imagine this similar to the organizational concept of the International Space Station.

*Openness* is an important organizational principle of the FuturICT. We envisage that the composition of the consortium of experts will continuously change over time to take new rising stars of science on board. Through open calls, FuturICT plans to allocate substantial research budgets to innovative research in order to reward excellent findings and support future research activities.

Openness will also be achieved by creating interfaces with business partners and policy-makers. In particular, the research infrastructures created by FuturICT (such as the Exploratories of Society, Economics, Technology, and Environment) will be open to researchers outside the FuturICT consortium. This concept and the multi-level structure of FuturICT are also designed for a steady expansion.

In fact, scalability is an important organizational principle of the project, as it is anticipated that there will be an increasing demand for research in the area represented by FuturICT, i.e. the research platform must be suited to support continuous growth and participation.

In order to support rapid scientific advances, research activities will be grouped around excellence clusters, i.e. a critical mass of experts in one institution, which is closely connected with the best international experts in other European locations. Each research core of FuturICT (see Figure 3) will jointly be led by three (or more) researchers stemming from different European regions and complementary knowledge areas. This will lead to a balanced leadership, which is operational at all times (by mutually backing up each other). These scientists will take strategic decisions in close interaction with a (variable) number of further experts contributing to the same research core and stay in close contact with the Committees overseeing and coordinating the FuturICT flagship. In fact, the Science Committee is composed of a subset of these core area leaders.

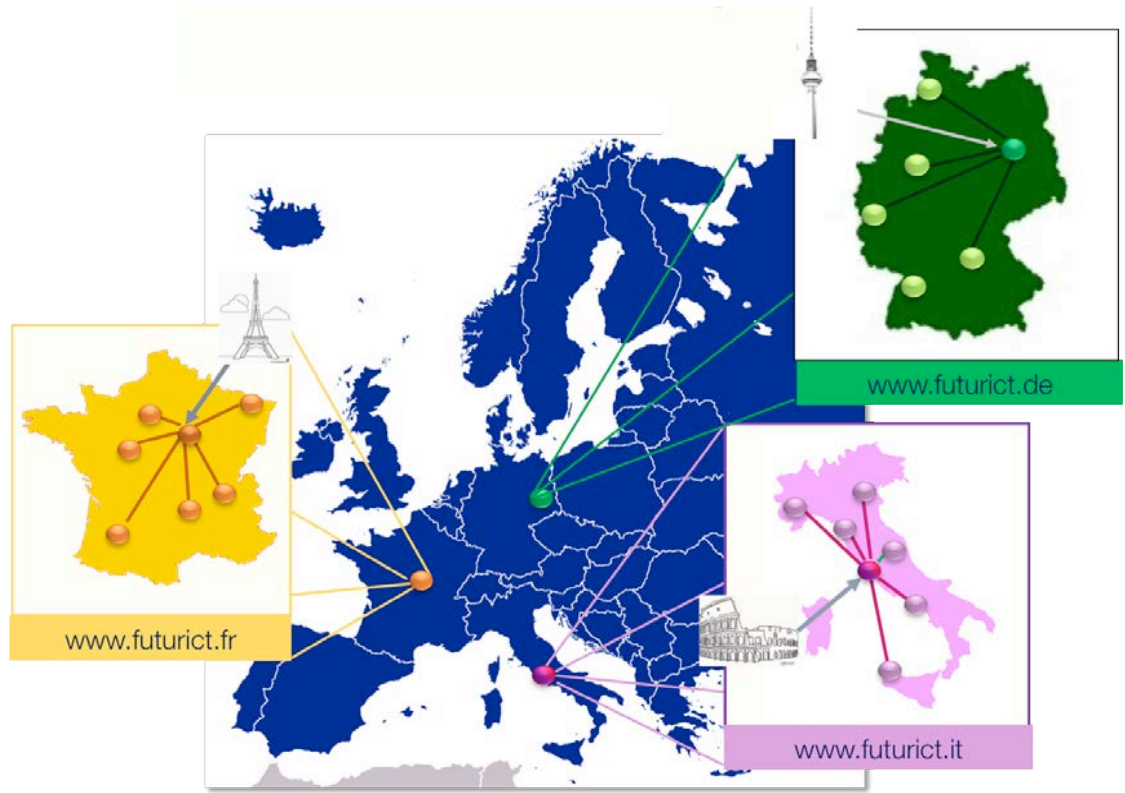

**Figure 7.** *FuturICT will be organized as a network of national hubs, with spokes to the various institutions in each country. Excellence centres, in contrast, are problem-specific international networks spanning several countries. This applies, in particular, to the Exploratories of Society, Economy, Technology and Environment. Consequently, there are two overlaid networks that help to integrate the communities and focus areas involved in the FuturICT flagship.*

FuturICT will have a *flexible and modern organization*, based on a multi-level hub-and-spoke network, integrating bottom-up elements. Besides the subject-level organization of FuturICT's research activities, FuturICT will build on strong national FuturICT communities, often integrating dozens of teams from the areas of ICT, complexity science, and the social sciences. These communities will help to stimulate and coordinate research activities beyond what the FuturICT project itself can fund, i.e. act as a Network of Excellence and Coordination Action. The national scientific communities will be involved

1. through annual national and international workshops (part of which will have a `Hilbert format', i.e. identify open problems and possible solution approaches rather than just presenting progress reports),

2. through awarding prizes to researchers for the best and most relevant results (providing money for follow-up research through the instrument of Open Calls).

The various academic institutions involved in FuturICT will receive their funds through one single national institution. Each national community will have a certain degree of *autonomy*, but also a set of obligations, e.g. to ensure agreed deliveries and reports. Responsibilities such as running jointly used Research Infrastructures or coordinating international Excellence Centres will be negotiated with the Steering Committee of FuturICT. All in all, the project may be roughly imagined as a super-large Integrated Project with one institution in each participating country as a partner, many subcontractors, and an integrated Coordination Action or Network of Excellence. European funds will be used to match national funds proportionally, where possible, and vice versa.

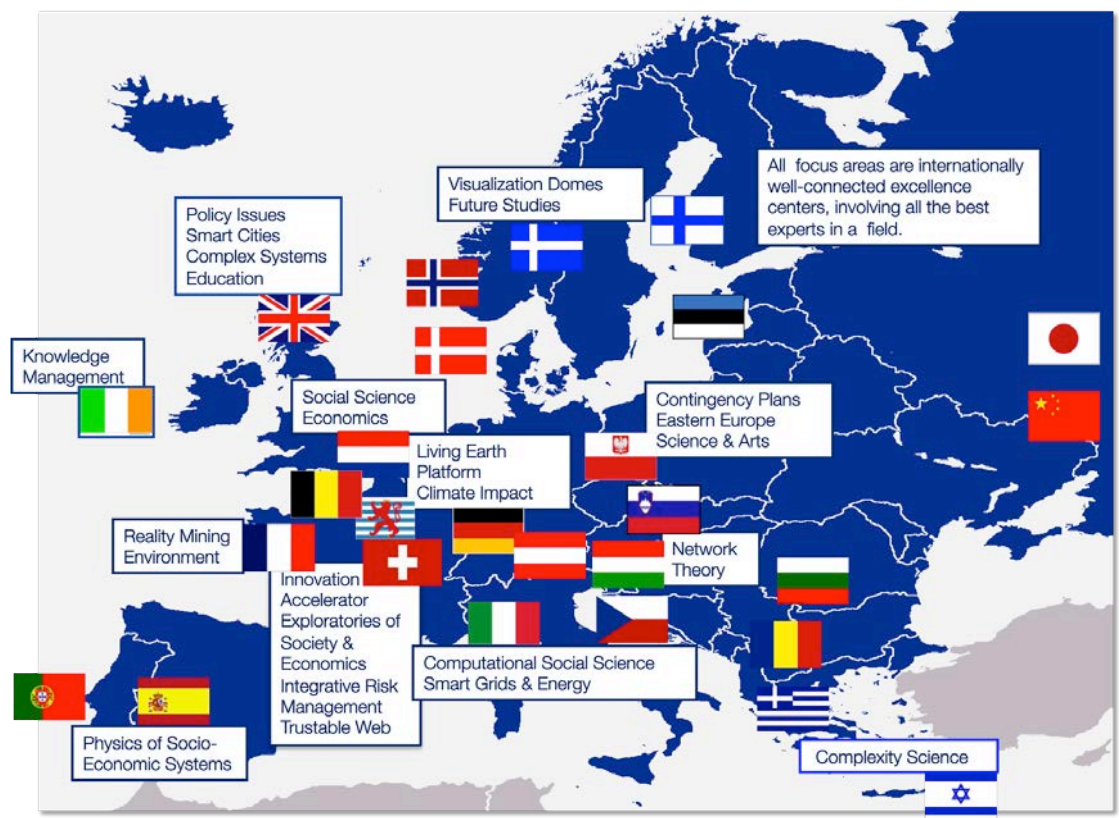

**Figure 8.** *Illustration of possible (non-exclusive) focus areas of the national FuturICT communities. All countries integrate activities in the areas of ICT, complexity and social sciences. The excellence centers will be internationally well connected and involve the best experts from all countries.*

**Summary and Conclusions**

FuturICT is not starting from scratch. It can stand on the shoulders of giants (see Boxes 16 and 17) and has learned from the experience of previous approaches (see Box 18). Building on several hundred teams of scientists in Europe and all over the world (see Box 15), FuturICT has a strong potential to promote the beneficial co-evolution of ICT and society, and also to encourage a new synthesis in and with the social sciences – supported by a plethora of computational methods for modelling, theory building, and analysis. Among other recent developments, the availability of Big Data will boost the development of the social sciences (see Box 14), and it will create a new information science. Beyond this, FuturICT will help overcome the current "data tragedy of the commons", which may be compared with a language in which everybody owns different words. Creating an open platform, a "data commons" (see Box 19) is expected to trigger off a new era of information and communication technologies, services, and products (see Figure 9 and Box 13).

The FuturICT Knowledge Accelerator will bring about a quantum leap in our capacity to more effectively cope with the speed at which our world is changing, and make a vital contribution to societal resilience and a sustainable future. It will do so by combining the best established scientific methods with multi-scale computer modelling, social supercomputing, large-scale data mining and participatory platforms (including web experiments and populated virtual worlds). Innovations needed to drive FuturICT forward to reach these ambitious goals will be promoted through a series of *'Hilbert Workshops'*, i.e. think tanks to identify the fundamental problems and how they can be solved. As a result, we expect to see a century of social and socio-inspired innovations, opening up new social, economic, and political opportunities. Indeed, there are many more reasons to make public investments into the FuturICT project (see Box 20).

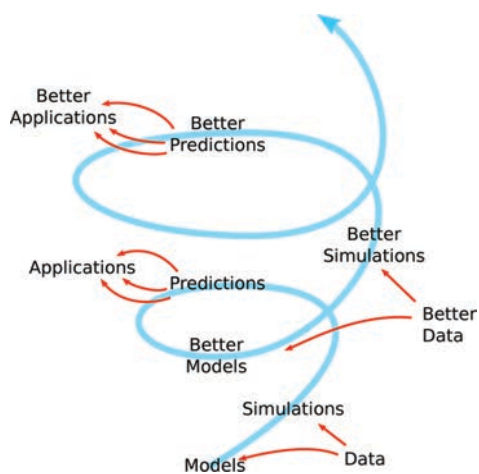

**Figure 9.** *Illustration of how Big Data of techno-socio-economic-environmental systems will boost better and better models, simulations, predictions, and applications. These will trigger off new business opportunities and spin-offs.*

**Further Information**

*Webpage* (including publications, media response, events, …):
http://www.futurict.eu

*Facebook* (including videos): http://www.facebook.com/FuturICT

*Twitter* (including interesting quotes): https://twitter.com/#!/FuturICT

*Visioneer* White Papers: http://springerlink.metapress.com/content/1951-6355/195/1/

**Acknowledgments**

The author would like to thank the FuturICT Advisory Board and the following people for valuable feedback: Steven Bishop, Giulia Bonelli, Mark Buchanan, Anna Carbone, Rosaria Conte, George Kampis, Kai Kunze, Paul Lukowicz, JB McCarthy, Suzy Moat, Andrzej Nowak, Tobias Preis, Rob Procter, Janet Smart, Hilary Woodard, and many others. The illustrations have been produced by Stefano Balietti, Steven Bishop, Anna Carbone, Dirk Helbing, George Kampis, Suzy Moat, and Pratik Mukerji.

# BOXES

**BOX 1: Paradigm Shift in Our Understanding of the World**

The conventional recipes to tackle the problems of our world fail more and more often. It seems that many established concepts are not fitting anymore. And applying more of the same medicine is usually not a good solution. Therefore, it has become a widely established view that we must learn to live with our current problems and deal with them like fire fighters.

However, many problems today are due to an out-dated understanding of our world. In fact, our traditional way of thinking is fundamentally wrong, because the world has changed: While its parts still look pretty much the same, we have networked them and made them strongly interdependent. This promotes an eigendynamics in the system. When such "self-organization" sets in, the components' individual properties are not anymore characteristic for the system behaviour, but collective behaviour takes over. Group dynamics and mass psychology are two typical examples.

Lee C. Bollinger, president of Columbia University, formulated the issue as follows: "The forces affecting societies around the world ... are powerful and novel. The spread of global market systems ... are ... reshaping our world ..., raising profound questions. These questions call for the kinds of analyses and understandings that academic institutions are uniquely capable of providing. Too many policy failures are fundamentally failures of knowledge."

As a consequence of the above, we have to turn our attention away from the visible components of our world to the invisible part of it: their interactions. In other words, we need a shift from an object-oriented to an interaction-oriented view, as it is at the heart of complexity science. This paradigm shift is perhaps of similar importance as the transition from a geocentric to a heliocentric worldview. It has fundamental implications for the way in which complex techno-socio-economic systems must be managed and, hence, for politics and economics. Focusing on the interactions in a system and the emergent dynamics resulting from them opens up fundamentally new solutions to long-standing problems (see Box 2).

**BOX 2: Some Success Stories: Applications of Recent Discoveries**

Fuelled by recent theoretical progress (see Box 14), the social and complexity sciences have sparked off a number of interesting use cases, particularly where both perspectives have been integrated: These use cases range from models of self-organization and segregation, suggesting strategies to reduce crime and conflict, over models of social cooperation, which imply recipes to overcome tragedies of the commons, up to opinion formation and wisdom of crowds models, which are used in so-called prediction markets. Models of pedestrian dynamics can now help to anticipate and avoid crowd disasters. Models of mobility patterns and traffic breakdowns support congestion avoidance and inform the design of smarter cities. Models of financial systems suggest, how to make them more stable and more resilient to shocks. Simulations of supply chains facilitate more efficient production systems and provide a better understanding of business cycles. Models of conflict and organized crime will allow one to reduce wars, insurgency, and drug traffic. It has also become possible to perform a real-time measurement and simulation of pandemics, which can be used for scenario-based policy recommendations, e.g. regarding more effective immunization strategies. Network theory allows one to understand and improve the resilience of systems. It has furthermore enabled Google's powerful page rank algorithm, and the semantic web as well as trust and recommender systems. In addition, social networking applications and models for the evolution of social groups and communities have mutually inspired each other. Last but not least, models for the emergence of language allow robots to find an own way of communication. All of this promises a great future and relevance of FuturICT's research.

**BOX 3: Ethical Issues**

FuturICT has a strong ethical motivation. In particular, the project wants to promote the development of responsive, responsible and ethical ICT.

Over a 10 years time period, FuturICT wants to provide an open data, simulation, exploration and participatory platform for everyone. This platform is thought to establish a new public good on which services of all kinds can be built, i.e. it will support both commercial and non-profit activities. To prevent misuse of the platform and enable reliable high-quality services, the platform will be built on principles of transparency, accountability, reputation, and self-regulation.

FuturICT is not interested in tracking individual behaviour or gathering data on individual actions. Its aim is to understand the macroscopic and statistical interdependencies within the highly complex systems on which we all depend.

The FuturICT project will have a strong research focus on ethical issues, and is committed to informing the public about the use of socio-economic data. For example, FuturICT will promote the development of a *Trustable Web* and of privacy-respecting data mining technologies that give users control over their data. It will strongly engage in preventing and counter-acting the misuse of data and the Internet. More broadly, the project will seek public involvement to build and sustain confidence in its values (see Box 4).

Finally, we consider it as a moral obligation to push the research directions promoted by the FuturICT project forward as quickly as possible. Given the fragility of the financial and economic system, the risks that this may finally impact the stability of our society and promote crime, corruption, violence, riots, and political extremism, or even endanger our democracies and our cultural heritage are not negligible anymore. Quick scientific progress is needed in order to learn how to efficiently stop the on-going cascading effects and downward trends. It is of similar importance to ensure that social and socio-inspired innovations will benefit humanity and not end up in the hands of a few stakeholders, as it partially happened in genetics (particularly food production).

**BOX 4: FuturICT's Values**

FuturICT wants to promote human well-being, increase the self-awareness of society, reduce vulnerability and risk, increase resilience (the ability to absorb societal, economic, or environmental shocks), reduce damages due to large-scale loss of control related to unexpected cascading effects and systemic shifts, develop contingency plans, explore options for future opportunities and challenges, increase sustainability, facilitate flexible adaptation, promote fairness and happiness, protect and increase social capital, support economic, political, and social participation, find a good balance between central and decentralised (global and local) control, protect privacy and other human rights, pluralism and socio-bio-diversity, and support collaborative forms of competition ("co-opetition").

**BOX 5: Resilient Self-Control and How to Make More Out of Scarce Resources**

Many crises result from domino or cascading effects (see Box 6). These may, for example, be compared with the formation of traffic jams. In fact, urban traffic flows form a strongly coupled system. The traffic flows from different points of origin towards different destinations may significantly influence each other. Classical traffic light control is based on supercomputing centres, which collect flow-rate measurements from many intersections and implement pre-determined control decisions in a top-down fashion. As the decision cannot be optimized in real-time (because there are too many alternative control options), one adapts a solution, which is optimal for, say, the typical Monday morning or the time after a soccer match. The average situation, however, never really occurs, since the variability of the flows from one red light phase to another is high. Hence, today's traffic light control is far from optimal. A better approach controls traffic flows bottom-up (or combines top-down and bottom-up elements in a suitable way), letting traffic lights flexibly respond to the actual traffic situation. When measuring not only the outflows from road sections, but also the inflows, a short-term anticipation of groups of vehicles becomes possible. This allows the traffic lights to turn green when a vehicle group arrives. It is interesting to note that a high-level of system performance is not reached, when each traffic intersection simply implements the best possible local control (as Adam Smith's principle may suggest). However, everyone profits if neighbouring intersections coordinate with each other through short-term flow anticipation: car drivers, users of public transport, bikers, pedestrians, and the environment as well. This example illustrates how complex, highly variable and largely unpredictable systems can be made more efficient and resilient: by a combination of real-time measurement, short-term anticipation, self-control, and the interruption of cascading effects (here: the avoidance of spill-over effects). As a result, crises can be prevented or mitigated, and scarce resources (in the above case: space and time) can be used in a better way.

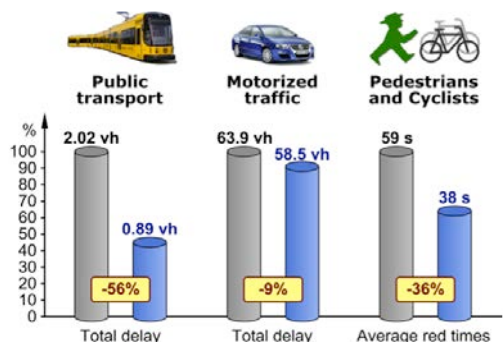

*In comparison with a state-of-the-art control (left columns), the flexible self-control (right columns) can reduce the average delay for all modes of transport and is beneficial for our environment as well.*

**BOX 6: Examples of Cascading Effects**

*In 2011 alone, 3 major cascading effects have occurred, which may change the face of the world and the global balance of power: The financial crisis, the Arab spring and the earthquake, tsunami and nuclear disaster in Japan. In the following, we discuss some examples of cascading effects in more detail.*

On November 4, 2006, an electricity line was temporarily turned off in Ems in Germany to facilitate the transfer of a Norwegian ship. This caused a chain-reaction leaving major parts of Europe without electricity. The foregoing scenario analysis had not checked for the coincidence with a possible spontaneous failure of another power line.

Heavy solar storms, as expected for the near future, could simultaneously bring down major parts of the worldwide ICT system, since most ICT systems are not sufficiently shielded against the related atmospheric currents of electrically charged particles. As a consequence, cash machines, sales and customer supplies, computer and communication systems could fail critically at the same time in large areas.

On December 22, 2010, Skype pushed a faulty auto-update of its Internet telephony software. This led to a crash and reboot of most Skype super-nodes, a crucial part of their distributed systems. To make things worse, the reboot of the super-nodes launched a distributed denial of service attack on the central Skype servers, thus incapacitating worldwide traffic.

The current economic crises started locally, due to a bursting real estate bubble in the US. The mortgage crises eventually hit building companies and caused the bankruptcy of more than 400 US banks. In the meantime, it endangers the stability of the European currency and even of the European Union. Several countries (including Greece, Ireland, Portugal, Spain, Italy and the US) are at the verge of bankruptcy. If the crises cannot be stopped, it will cause social unrest, political extremism and increasing crime and violence, which have the potential to endanger the cultural foundations of our society and peace.

The 2011 Tōhoku earthquake in Japan caused a tsunami that triggered chain reactions and nuclear disasters in several reactors at Fukushima. Soon after, several countries, including Germany and Switzerland, decided to exit nuclear energy generation over the next decade(s). However, alternative energy scenarios turn out to be politically vulnerable. Two of three major regions currently providing Europe with gas may not be entirely reliable. Moreover, Europe's DESERTEC project, which planned to invest 1000 billion EUR into infrastructure to supply solar energy for Europe, has now an uncertain future due to another political unexpected event, the Arab Spring. This was triggered in particular by high food prices, which were no longer affordable to many people. These high food prices resulted partly from biofuel production, which intended to improve the global $CO_2$ balance, but competed with food production. The increasing food prices were further amplified by financial speculation.

The spreading of flu outbreaks is very much promoted by worldwide travel and sometimes food supply chains as well. In the case of the resulting pandemics, economic and social life can be enormously affected (see Box 12).

**BOX 7: Research Priorities**

FuturICT will build its global ICT Platform by integrating various Observatories into interconnected Exploratories, which will eventually be fused to form a single, integrated infrastructure (see Figure 6). The interconnected Observatories and Exploratories will feature real-time data-mining, computer simulations to facilitate scenario analyses, and interactive environments allowing people to explore realistic virtual worlds. As the Observatories start off from already existing cores, first practically relevant applications are expected after 2 years. Practical use cases of the FuturICT flagship include smart cities (as more than 50% of the world population is now living in cities), smart energy systems (as micro-generation of electrical power will increase the number of independent energy providers multiple times), smart health systems (to ensure a high quality of life in an ageing society at affordable costs), and better financial architectures (to reduce the related societal losses). New solutions will also be developed to reduce crime, corruption, and conflict. Eventually, FuturICT's Participatory Platform will inform decision-makers and involve citizens. A focus on *Managing Complexity* will develop integrative system designs and new decision-making and governance tools. The *Innovation Accelerator* will speed up research, development, and the creation of new business opportunities. Finally, a broad spectrum of socially inspired ICT will be developed.

To gain a better understanding of social, economic and ICT systems, a number of over-arching research challenges must be addressed, such as: 1. the interrelationship between structure, dynamics, and function, 2. strongly coupled systems and interdependent networks, 3. contagion and cascading effects, 4. ecological and social systems thinking, 5. managing complexity, 6. incentives, 7. integrative systems design, 8. resilience vs. systemic risks, 9. sustainability, and 10. trust.

Over a ten years time period, FuturICT is expected to fill the current knowledge gaps and to greatly advance the following new research areas: non-equilibrium economics, the science of strongly coupled and interdependent systems, the science of multi-level complex systems, a new data science, and methods to assess systemic risks and manage them in an integrated way. Presently, these areas appear like continents just discovered, and many new breakthroughs are expected almost as it was possible at Humboldt's times.

**BOX 8: The European Flagship Program and its Call for Big Science**

The FuturICT project is the response to a call from the Future and Emerging Technology (FET) section of the European Commission, hence the name FET Flagship. The objective is to support Big Science in Europe with a "Man on the Moon" type vision. In the first round, 21 flagship candidates were narrowed down to six Flagship Pilots. FuturICT was determined as leading Flagship Pilot proposal, addressing techno-socio-economic and environmental challenges of the future. A comparison with the other Flagship Pilots is offered by Box 9.

Each pilot will submit detailed proposals in April 2012. At least two flagship projects will be supported by an amount of up to 1 billion EUR each over a time period of 10 years. This is about a tenth (or less) of what is invested into other Big Science projects: the CERN elementary particle accelerator, the ITER fusion reactor, or the Galileo satellite program, the Human Genome project, nanotechnology, etc. Approximately half of the money, i.e. 50 million EUR per year, must be mobilized by the project partners from national budgets and funding agencies, from business and industry, or from donations. It is planned to distribute a considerable fraction of the flagship budget through Open Calls. This will allow a wide scientific community to contribute to the common goals of the project.

**BOX 9: What Distinguishes FuturICT from Other Candidate Flagships**

- FuturICT will develop a new science of big data, allowing one to understand how an ocean of information bits can be turned into useful knowledge, wisdom, and business opportunities.
- FuturICT will build a Planetary Nervous System by harvesting the data streams from smart sensors that are now becoming commercially available and spreading all over the world.
- FuturICT aims at building up non-embodied artificial intelligence and connecting the brains of millions or even billions of people to promote creativity and collective intelligence.
- FuturICT is promoting global health by building a related Observatory and identifying social ways of spreading healthy behaviour.
- FuturICT will promote social well-being. As humans prefer to be helped by others, the project also seeks ways to promote the mutual understanding and solidarity in increasingly multi-cultural and fragmented societies.

The particular strengths of the FuturICT project are its particular societal relevance, the immediate importance of its results to master our everyday life in the future, its large and quickly growing community integrating multiple disciplinary backgrounds, the participation of multiple European countries, the significant support of scientific communities in other continents, the strong focus on ethical issues, the open project architecture, the innovative organizational concept with its bottom-up elements, and the remarkable activities in the area of education. Further benefits of FuturICT are discussed in Box 20.

**BOX 10: How to Establish a Planetary Nervous System**

The goal of creating a planetary nervous system is to measure the state of the world and the interactions in it. For this, real-time data mining, so-called "reality mining", will be established, using data of the Internet and the semantic web. Additionally, data will be collected by linking sensors which aggregate information about the technological, social, or economic activities around them. Such a global sensor network can, for example, be established by connecting the sensors in today's smartphones (which comprise accelerometers, microphones, video functions, compasses, GPS, and more). Here, FuturICT will closely collaborate with Prof. Sandy Pentland's team at MIT's Media Lab.

In order to reach that users will contribute own data voluntarily, a number of criteria must be fulfilled: 1. The system must provide incentives (such as sharing benefits and profits generated with the provided data). 2. A micropayment system is needed to establish a market for the reward points earned. 3. Users must be given control over their own data and what they are used for. 4. A privacy-respecting data mining approach must be developed. 5. Macroscopic measurement procedures must be invented to anonymize and aggregate sensitive data "on the fly".

Two illustrative examples for smart-phone-based collective sensing applications are the open streetmap project and a decentralized earthquake sensing and warning concept.

Sandy Pentland: Society's Nervous System: Building Effective Government, Energy, and Public Health Systems, http://hd.media.mit.edu/tech-reports/TR-664.pdf

**BOX 11: Possibilities and Limits of Prediction**

FuturICT is often confronted with questions regarding the predictability of its models. Recent findings suggest that the dynamics in social systems depends at least on three different factors: the situational context, interaction effects, and random events. Therefore, the measurement of the first two factors should, in principle, allow for probabilistic forecasts.

In fact, a systemic analysis, combined with a situational analysis, allowed one to predict developments such as the destabilization of the financial system or the overall political impact of September 11 on various countries. Nevertheless, long-term forecasts are not a goal of FuturICT. We recognize that long-term forecasts are restricted to a few global trends (such as Moore's law or population growth and related resource issues), and that forecasting the exact timing is generally difficult in social systems. Short-term anticipation, however, is often sufficient to reach significant improvements by adaptive strategies (e.g. principles of self-organization and self-control, see Box 5).

In addition, it is often possible to predict likely courses of events, since cascading effects follow from causal relationships (see Figure below). Note that an assessment of the robustness of a system does not require *forecasting* (i.e. *when* something will be happening), but only a *predictive model* (that says *under what conditions* something is likely to happen). Given suitable data, it increasingly becomes possible to determine how the state of a complex system depends on the properties of the system components, their interactions, the environment, institutional setting, and resources. Given the availability of situational and contextual data, it should also become possible to determine the impact of predictions on social systems, namely whether the case of a self-defeating prophecy is expected to occur, or the case of a self-fulfilling prophecy, or the case where the prediction has no significant effect at all. Finally, as has been experimentally demonstrated, it is possible to design recommender systems in such a way that their usefulness is not undermined by information feedbacks.

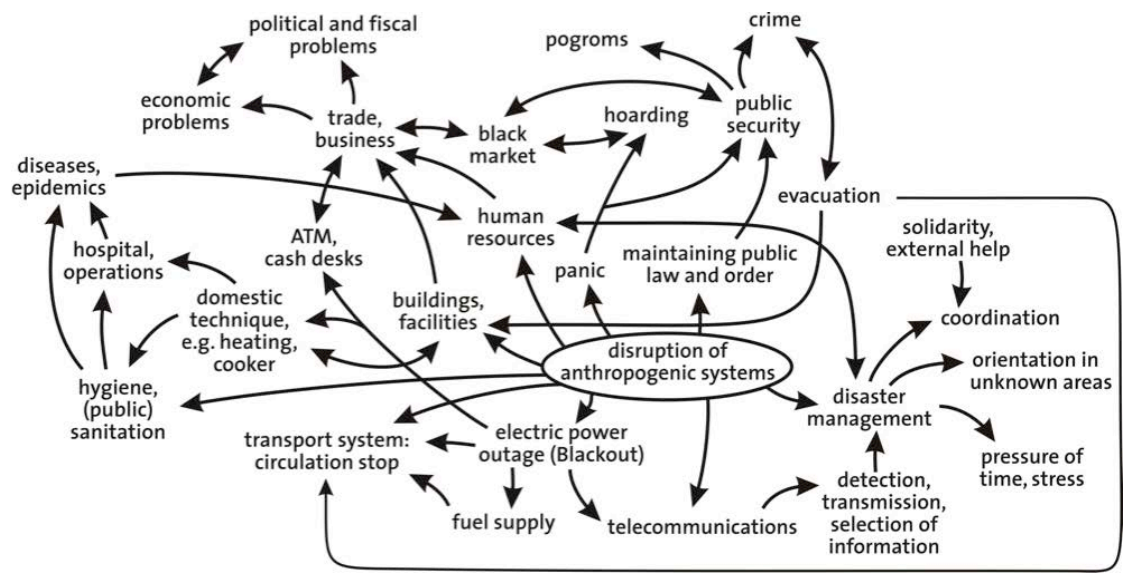

*Illustration of cascading effects in techno-socio-economic-environmental systems, which may be triggered by the disruption (over-critical perturbation) of an anthropogenic system. A more detailed picture can be given for specific disasters. Note that the largest financial damage of most disasters is caused by such cascading effects.*

**BOX 12: Costs and Benefits of the FuturICT Project**

Among other challenges, FuturICT will develop new concepts to address the following problems, the societal costs of which are listed below:

1. *Financial crisis:* Losses of 2.2 trillion $
2. *Crime and corruption:* 2-5% of GDP, about 2 trillion $ annually.
3. *Conflict:* Global military expenditures of 1.5 trillion $ annually.
4. *Terrorism:* 90 billion $ lost output of the US economy as a result of 9/11 attacks.
5. *Flu:* A true influenza pandemic infecting 1% of the world population would cause losses of 1-2 trillion $ per annum.
6. *Congestion:* Impact of 7-8 billion £ in the UK alone.

If the impact of these societal problems would be reduced by 1% only, this would already create a benefit many times higher than the investments into the FuturICT Flagship project. Based on previous success stories regarding a better management of complex systems, we expect that an improved understanding of the fundamental underlying issues will facilitate improvements between 10% and 30%, given the new insights are properly applied. For comparison: Swiss citizens pay 10 Swiss Francs per year for the public weather forecasting system, but the benefit is 5 times higher.

*Benefits expected from ICT-related growth and productivity:* ICT-producing industries contribute directly to productivity and growth through their own rapid technological progress. For example, a rough estimate indicates that in the United States in 2008, Internet intermediaries contributed at least 1.4% of GDP value added. This produces 'spill-over effects' on the rest of the economy as ICT diffusion leads to innovation and efficiency gains in other sectors. A good example for the economic potential of socio-inspired ICT is the company Facebook, the value of which is believed to have exceeded already 65 billion $.

Further prospects and benefits of the FuturICT project are summarized in Box 20.

**BOX 13: FuturICT's New Approach to ICT Systems and Innovation**

Humans are a unique species, as their behaviour is largely driven by information. By the creation of virtual worlds and in many other ways, future ICT systems will partly overcome the limitations of our physical and biological world. In fact, they will create an almost unlimited number of new goods and services, and thereby many new economic opportunities, but also social and political ones.

Europe's vision of creating an Innovation Union is a logical response to these opportunities. With its *Innovation Accelerator* approach, FuturICT will significantly contribute to laying the foundations of this Innovation Union. Through new ICT technologies, *Innovation Scouts* and *Knowledge Transfer Supply Chains*, the distance between academic inventions and innovations in the technological, social and political realm will be significantly reduced. In this way, transforming new ideas into new products will be much more efficient than today (currently this requires of the order of 30 years in many areas).

Fundamentally new ICT systems that are responsive, responsible, ethical and privacy-respecting by design, are key to exploring, understanding, and managing our future in a resilient and sustainable way. The FuturICT flagship will promote the required paradigm shifts

- by integrating the best knowledge from the engineering, natural and social sciences, and bringing together Big Data, explanatory theories, massive computer simulations, and large-scale experiments in virtual and real-world settings, and
- by revolutionizing the social sciences and future ICT systems, asking the right questions and employing a complex systems perspective to understand the interaction-based and institutional principles that make strongly connected, socially interactive systems work well.

This will trigger a new era of the social sciences and a wave of socio-inspired technologies (beyond social networking, the wisdom of crowds, and prediction markets). Future ICT systems will have the capability of social sensing, social thinking, social adaptiveness, social self-organization, etc. They will feature ICT-based cultures, collective (self-)awareness, reputation- and trust-based applications (such as "social money") etc. And they will feature mixed reality systems, where it won't be possible to tell apart the virtual and real world. The goal will be a beneficial *human-information symbiosis*.

The creation of such an *information ecosystem* will require a new kind of complexity science, which is capable of realistically understanding multi-level complex techno-socio-economic systems. This new complexity science will, in turn, facilitate the reduction of systemic risks by employing suitable decoupling strategies and the design of resilient and sustainable socio-economic and ICT systems. However, a well-functioning information ecology also needs to overcome the lack of transparency, accountability, quality standards and trustworthiness of most current data services provided on the web for free. Currently, many companies collect huge data sets, but these are often fragmented and potentially sensitive.

As the World Economic Forum points out, users must be given control over their personal data [http://www3.weforum.org/docs/WEF_ITTC_PersonalDataNewAsset_Report_2011.pdf]. Moreover, it is important to develop methods of privacy-respecting data mining, which can satisfy individual, commercial and public interests at the same time [for a proposal how to do this, see Dirk Helbing and Stefano Balietti: "From Social Data Mining to Forecasting Socio-Economic Crisis", http://arxiv.org/abs/1012.0178]. This will require a scalable bottom-up approach, transparency, user control, encryption of sensitive data and digital rights management, plus a manipulation-proof reputation system that supports a healthy "immune response" to malicious data and activities. In this way, it will be possible to create a self-controlled, and responsible future Internet (a *Trustable Web*). The creation of an integrated multi-disciplinary self-organized reputation-based science platform and the invention of a reputation system helping to avoid tragedies of the commons in a globalizing world will be two important use cases of this new Web.

**BOX 14: Big Data: A New Era of the Social Sciences is Ahead**

In the past, getting data about social systems and social interactions was very time consuming and cumbersome. In the meantime, lab and web experiments and online surveys have simplified the collection of data, and the internet as well as other information and communication systems are collecting tons of data allowing one to study social activity patterns. This is opening up the door for a new era of the social sciences. In parallel, much progress has been made in modelling key elements of social systems. Now, there are models considering spatial and network interactions, heterogeneity, and randomness (which can change the systemic outcome dramatically!). There are also models of emergence of cooperation under unexpected conditions (namely in social dilemma situations, which normally promote a 'tragedy of the commons'), models for the formation of social norms (even when individuals have to make sacrifices for this), and models for the spreading of conflicts or violence, models of collective behaviour (such as opinion formation, crowd disasters, revolutions), as well as models taking into account communication and learning. Currently, scientists are working on models considering emotions, models explaining conditions for other-regarding behaviour, and models considering cognitive complexity. All these models imply useful applications that are beneficial for society (see Box 2), given the availability of Big Data to calibrate and validate them.

**Box 15: FuturICT's Partners and Supporters**

By now, FuturICT is supported by more than 700 scientists worldwide. It involves Europe's academic powerhouses, such as ETH Zurich, University College London (UCL), Oxford University, the Fraunhofer Society, the Consiglio Nazionale delle Ricerche (CNR), the Centre National de la Recherché Scientifique (CNRS), Imperial College, etc. Five supercomputing centers support FuturICT, including the national ones in Germany, Switzerland, and Spain. There are also letters of support by the OECD, the Joint Research Center (JRC), regulatory authorities, international companies, and notable individuals such as George Soros. Furthermore, FuturICT has managed to integrate many different research communities, as the pie chart of Figure 2 illustrates. In fact, the leaders of FuturICT have a long track record of successfully integrating scientists across disciplinary boundaries, as is reflected by dozens of multi-disciplinary workshops in the past years. FuturICT's supporters have also been involved in hundreds of successful projects with business partners.

**BOX 16: History**

A strong historical backdrop is provided by the Digital Earth project, see http://www.digitalearth-isde.org/ and http://en.wikipedia.org/wiki/Digital_Earth.

The following quotes are from the above Wikipedia page, accessed on July 24, 2011:

"In a speech prepared for the California Science Center in Los Angeles on January 31, 1998, [the former US vice president Al] Gore described a digital future where schoolchildren - indeed all the world's citizens - could interact with a computer-generated three-dimensional spinning virtual globe and access vast amounts of scientific and cultural information to help them understand the Earth and its human activities…

Digital Earth has come to stand for the large and growing set of web-based geographic computing systems worldwide. These are both useful and promising, but do not yet constitute the envisioned `global commons'."

Below follow two excerpts from the Beijing Declaration on Digital Earth
[http://159.226.224.4/isde6en/hykx11.html]:

"Digital Earth is an integral part of other advanced technologies including: earth observation, geo-information systems, global positioning systems, communication networks, sensor webs, electromagnetic identifiers, virtual reality, grid computation, etc. It is seen as a global strategic contributor to scientific and technological developments, and will be a catalyst in finding solutions to international scientific and societal issues."

"Digital Earth should play a strategic and sustainable role in addressing such challenges to human society as natural resource depletion, food and water insecurity, energy shortages, environmental degradation, natural disasters response, population explosion, and, in particular, global climate change."

Considering this, FuturICT may be seen as a logical continuation of the Digital Earth Agenda with a focus on
1. exploring and managing socially interactive systems,
2. real-time mining and modelling of techno-socio-economic data to promote collective (self-)awareness, and
3. creating participatory platforms including populated virtual worlds.

**BOX 17: Reference Cases**

Selected reference cases further illustrating FuturICT's feasibility include:

- *IBM SmarterPlanet* (http://www.ibm.com/smarterplanet),
- *Microsoft Modeling the World* (http://www.modelingtheworld.com/)
- *Earth Simulator (Japan)* (http://www.jamstec.go.jp/esc/index.en.html, http://en.wikipedia.org/wiki/Earth_Simulator)
- *PAX Early Warning of Conflict* (http://www.pax2011.org/index.php)
- *Sentient World White Paper* (http://www.scribd.com/doc/25656152/Sentient-World-Simulation)
- *Planetary Skin Institute* (http://www.planetaryskin.org/)
- *Second Life* (http://secondlife.com/)
- *Google.org* (http://www.google.org/)
- *Gapminder* (http://www.gapminder.org/)
- *Observatorium* (http://www.observatorium.eu)

**BOX 18: How FuturICT Differs from Previous Approaches**

There have obviously been previous attempts to model the dynamics of the world. FuturICT has learned from them. It has a more sophisticated and more differentiated approach and is supported by hundreds of scientists worldwide. FuturICT is not aiming at long-term forecasts. It addresses current and generic problems. It aims at improving the system performance and the ability to absorb shocks (see Box 5 for its other goals). FuturICT's models will consider spatial and network effects, heterogeneity and randomness. They will build on the availability of Big Data and the possibility of real-time data mining as well as the progress in network theory, complex systems theory, multi-agent simulation, multi-level modelling, computational social science, experimental approaches and interactive platforms. Furthermore, FuturICT will develop new methods of investigation such as a interconnected Exploratories, Living Earth Simulator, a Planetary Nervous System, a Global Participatory Platform.

**BOX 19: Creating an Open, Transparent Platform for Everyone**

FuturICT wants to overcome the current data fragmentation and "black holes" for data. Instead, it is trying to create an open platform for everyone. This includes to establish transparency regarding the data sources and their quality, the exact algorithms used, the statistical assessment of the results. Furthermore, it will be important to establish transparent, responsible use – a subject worked on by FuturICTs ethics committee. The result will be a new public good, like our environment, air, languages, and the Internet. This will enable an ecosystem of new services and jobs, and an age of creativity. The goal is to remove barriers for social, economic, and political participation. However, a public good requires measures to prevent a 'tragedy of the commons', such as data pollution, manipulation, misuse, and cybercrime.

In order to build a Trustable Web, one needs to ensure control of users over their own data and the way they are used (see Box 13). One needs to create privacy-respecting information systems (and, hence, the use of pseudonyms). To promote responsible behaviour, a decentralized, transparent, and manipulation-resistant reputation system for information providers and contents is needed, together with a proper incentive system. This will establish a self-organizing and self-regulating system, something like a socially adaptive and mutually beneficial information ecosystem. To design the system properly, we need to understand socially interactive systems, otherwise one will end up with the same problems as in our society, i.e. instabilities, coordination failures, poor system performance and tragedies of the commons, conflicts, (cyber-)crime, and (cyber-)wars.

**BOX 20: Some Reasons to Publicly Invest into the FuturICT Flagship Project**

1. Considering what is at stake (see Boxes 3 and 12), there is a moral responsibility to do what we *can* do to address the 21st century challenges.
2. A federated Big Science approach is needed to catch up with the pace at which the world is changing and new political, social, economic and technological problems are emerging.
3. FuturICT will provide policy- and decision-makers with innovative methods and instruments to improve the societal, economic, and political situation, whereas commercial companies have no such capacity.
4. FuturICT will ultimately create tools to tackle social and natural catastrophes on a large scale. Individual private companies do not provide such tools.
5. FuturICT is an ethically oriented project and builds bridges between many scientific communities, which have previously worked in separation.
6. FuturICT will be a major driving force for *all* scientific research in the areas of ICT, social sciences and complexity science. It is triggering off entirely new trends in research and development, even beyond the research activities funded by the project. FuturICT will publish its results to benefit everyone, while private companies tend to keep their data, methods, and results for themselves.
7. FuturICT pursues an open platform approach under European leadership, which will allow other countries to participate (e.g. Japan, China, Singapore, Australia, South America, Africa).
8. FuturICT supports cooperative behavior on a global scale, while companies mostly tend to engage in competition.
9. FuturICT creates outcomes that profit-driven companies are unlikely to produce (such as privacy-respecting data mining technologies, a Trustable Web, or a public data platform providing a high-quality common good).
10. It must be avoided that powerful tools and social innovations end up in the hands of a few stakeholders rather than benefiting humanity.

---

[a] In the context of this document, the word "will" assumes that enough resources will be available to the project and that institutional constraints do not obstruct its progress or change its goals or roadmap. The document reflects what the FuturICT partners and many of its supporters currently think *should* be done.